\documentclass[preprint, pt1]{elsarticle}

\usepackage[latin1]{inputenc}                    
\usepackage{graphicx}                            
\usepackage{latexsym}                            
\usepackage{amsfonts}                            
\usepackage{amssymb}                             
\usepackage{amsmath}                             
\usepackage[mathscr]{eucal}                      
\usepackage{dcolumn}                             
\usepackage{theorem}                             
\usepackage{footnote}
\usepackage{enumitem}
\usepackage{bm}
\usepackage{color}
\usepackage{natbib}
\usepackage[margin=2.5cm]{geometry}%

\def\be{\begin{equation}}
\def\ee{\end{equation}}
\def\bea{\begin{eqnarray}}
\def\eea{\end{eqnarray}}

\def\gg{\gamma \gamma}
\def\gZ{\gamma Z}

\def\qwl{$Q_W^e \,$}

\def\qwe{$Q_{\text{weak}} \,\,$}
\def\mo{MOLLER }
\def\so{SoLID }

\def\bgZv{$\square_{\gZ}^V \,$}
\def\bgZa{$\square_{\gZ}^A \,$}
\def\regzv{$\Re e \, \square_{\gZ}^V$}		

\def\ftgz{F_2^{\gZ}}
\def\fogz{F_1^{\gZ}}

\def\g2{GeV$^2$}

\def\gev{\,{\rm GeV}}

\begin{document}

\title{Hadronic $\gamma Z$ box corrections in M\o ller scattering}

\author[adl]{N.~L.~Hall}

\author[man]{P.~G.~Blunden}

\author[jlb]{W.~Melnitchouk}

\author[adl]{A.~W.~Thomas}

\author[adl]{R.~D.~Young}

\address[adl]{\mbox{ARC Centre of Excellence for Particle Physics 
	at the Terascale and CSSM}, School of Chemistry and Physics,\\ 
	University of Adelaide, Adelaide SA 5005, Australia}

\address[man]{\mbox{Department of Physics and Astronomy,
	University of Manitoba}, Winnipeg,  MB, Canada R3T 2N2}

\address[jlb]{\mbox{Jefferson Lab, 
	12000 Jefferson Avenue, Newport News, Virginia 23606, USA}}

\date{\today}

\begin{abstract}
The possibility of measuring the parity-violating asymmetry
in M\o ller scattering with sufficient accuracy to determine
$\sin^2\theta_W$ to 0.1\% offers a complementary path to the
discovery of new physics to that followed at high energy colliders.
We present a new calculation of the $\gZ$ box contribution to
parity-violating electron-proton scattering, which constitutes an
important uncertainty in computing the background to this process.
We show that while the $\gZ$ correction grows rapidly with energy,
it can be relatively well constrained by data from parity-violating
inelastic scattering and parton distribution functions.
\end{abstract}


\maketitle

\section{Introduction}

In the search for physics beyond the Standard Model, low-energy 
precision tests of parity violation (PV) provide crucial alternatives
to searches at colliders such as those at the Large Hadron Collider.
Following pioneering work at SLAC \cite{Prescott:1979dh} and
MIT-Bates \cite{Souder:1990ia}, Jefferson Lab has recently seen
several such experiments.  The first, involving elastic
electron--proton scattering~\cite{Aniol:2000at, Aniol:2005zg,
Armstrong:2005hs}, led to the determination of the strangeness
contribution to the nucleon electromagnetic form factors
\cite{Paschke:2011zz, Young:2006jc}, as well as significant
new limits on the quark weak couplings $C_{1u}$ and $C_{1d}$
\cite{Young:2007zs,Roche:2011zz}.
More recently, the first report from the \qwe experiment
\cite{Qweak13} significantly improved those limits,
while a final, higher precision result is expected soon.
The most recent experiment involved the measurement of inelastic 
electron--deuteron scattering in the nucleon resonance region
and beyond \cite{Wang:2013kkc}.
All of these experiments were completed using Jefferson Lab's
6~GeV polarised electron beam and served to test the running of
$\sin^2\theta_W$ at low $Q^2$.  Following the 12~GeV upgrade
of the CEBAF accelerator, a new generation of PV experiments,
such as \mo \cite{Mammei:2012ph} and \so \cite{Souder:2012zz},
will provide even more stringent constraints on the Standard Model.

In this Letter we examine the M\o ller scattering process, and
in particular the $\gZ$ radiative corrections to the background
$e\, p$ scattering, which presents one of the main theoretical
uncertainties to this process.  The \mo experiment will scatter
longitudinally polarised electrons from atomic electrons in a
liquid hydrogen target with the aim of measuring the weak charge
of the electron, \qwl, to within 2.3\%~\cite{Mammei:2012ph}.
This will be equivalent to determining $\sin^2\theta_W$ to
$\approx 0.1\%$, placing it on par with the two (different)
values for $\sin^2\theta_W$ measured at the $Z$ pole.
This is especially important since these two values differ by
3$\sigma$, and, although their average is consistent with other
experimental data, if either of them were found to be the correct
value, the behaviour of $\sin^2\theta_W$ would change markedly
\cite{Mammei:2012ph} from that expected within the Standard Model.
Even if the average were indeed correct at the $Z$ boson pole,
a measurement to this precision would provide important information
on the nature of possible new physics~\cite{Mammei:2012ph, Erler:2003yk,
Erler:2013xha}.

The PV asymmetry in M\o ller scattering is defined as
\be
A_{\rm PV} = \frac{\sigma_+ - \sigma_-}{\sigma_+ + \sigma_-},
\ee
where $\sigma_\lambda$ is the cross section for an incoming
right-handed (helicity $\lambda = +1$) or left-handed (helicity
$\lambda = -1$) electron.  At the kinematics relevant to the \mo
experiment, the asymmetry is dominated by the interference between
the tree-level $\gamma$ and $Z$ exchanges, and is given by
\cite{Derman:1979zc},
\be
A_{\rm PV}
= m_e E\,
  \frac{G_F}{\sqrt{2} \pi \alpha}
  \frac{2y(1 - y)}{1 + y^4 + (1 - y)^4}\, Q_W^e,
\label{eq:apvdisd}
\ee
where $m_e$ and $E$ are the incident electron's mass and energy,
respectively, $y$ is the fractional energy transferred, $G_F$ is
the Fermi constant and $\alpha$ the fine structure constant.
At tree (or Born) level the weak charge of the electron is
given by $Q_W^{e \rm (Born)} = -1 + 4\sin^2\theta_W$.
For a determination of \qwl to a precision of $\approx 2.3\%$,
higher order radiative corrections must also be included.
These have been calculated in Refs.~\cite{Czarnecki:1995fw,
Denner:1998um, Petriello:2002wk} using standard techniques.

Because the \mo experiment uses a hydrogen target, the measurement of
the PV asymmetry unavoidably includes a background contribution from PV
$ep$ scattering, which depends on the weak charge of the proton, $Q_W^p$.
The \qwe experiment should determine the effective (or energy-dependent)
proton weak charge to an accuracy of 4\% at an energy of 1.165~GeV, where
the overall radiative corrections shift the Born result by around 75\%.
Of particular importance is the $\gZ$ box contribution associated with
the vector coupling of the $Z$ boson at the proton (axial-vector coupling
at the electron), \bgZv.  While this constitutes a modest, $\approx 7\%$
correction to $Q_W^p$ at the \qwe energy, it initially grows linearly
with energy and is found to be significantly more important
($\approx 15\%$) at the \mo energy of $E=11$~GeV.

In fact, there has been considerable interest recently in accurately
computing the absolute value and uncertainty of the \bgZv correction
\cite{Gorchtein:2008px, Sibirtsev:2010zg, Rislow:2010vi,
Gorchtein:2011mz, Blunden:2011rd, Hall:2013hta}, particularly how
it impacts the interpretation of the \qwe experiment.
The most recent analysis \cite{Hall:2013hta} used the
Adelaide-Jefferson Lab-Manitoba (AJM) model \cite{Hall:2013hta}
to provide a precise determination of the $\gZ$ correction,
utilising constraints from PV $ed$ inelastic scattering data
\cite{Wang:2013kkc} and global parton distribution functions (PDFs).
Since it is essential for the \mo experiment that the total
error on the proton weak charge at 11~GeV be less than 4\%,
it is necessary to ensure that the \bgZv correction here also
is under control.
Unfortunately, to date there has been no estimate of this
correction at the kinematics relevant to the \mo experiment.
We do so in this Letter.

\section{Adelaide-Jefferson Lab-Manitoba model}
\label{sec:AJM}

In presenting our calculation of the $\gZ$ correction to the inelastic
background for the \mo experiment, we begin by briefly summarising the
salient features of the AJM model as relevant for the present analysis;
further details can be found in Ref.~\cite{Hall:2013hta}.
At tree level, the weak charge of the proton is given by
  $Q_W^{p {\rm (Born)}} = 1 - 4\sin^2\theta_W$,
while at higher orders additional radiative corrections are important,
the most challenging of which is the $\gZ$ interference box term, \bgZv.
From the crossing symmetry properties of the vector hadron part of the
$\gZ$ correction, the real part of \bgZv can be written with the help
of a forward dispersion relation in terms of its imaginary part
\cite{Gorchtein:2008px},
\be
\Re e\, \square_{\gZ}^V (E)\
=\ \frac{2E}{\pi}\,
  {\cal P} \int_0^\infty dE' \frac{1}{E'^2-E^2}\,
  \Im m\, \square_{\gZ}^V(E'),
\label{eq:DR}
\ee
where ${\cal P}$ denotes the principal value integral.
Using the optical theorem, the imaginary part of \bgZv can be computed
from the vector interference $\gZ$ structure functions $F_{1,2}^{\gZ}$
as \cite{Gorchtein:2008px, Sibirtsev:2010zg, Arrington:2011dn}
\be
\Im m\, \square_{\gZ}^V(E)\
=\ {1 \over (s - M^2)^2}
  \int_{W_\pi^2}^s dW^2
  \int_0^{Q^2_{\rm max}} dQ^2\, {\alpha(Q^2) \over 1+Q^2/M_Z^2}
  \left[ \fogz
         + { x s \left( Q^2_{\rm max}-Q^2 \right) \over Q^4 } \ftgz
  \right],
\label{eq:ImBoxV}					
\ee
where $Q^2$ is (minus) the squared mass of the exchanged virtual
$\gamma$ or $Z$ boson, $W$ is the invariant mass of the hadronic
intermediate state, and $x = Q^2/(W^2 - M^2 + Q^2)$ is the Bjorken
scaling variable, with $M$ the nucleon mass, and $M_Z$ the mass of
the $Z$ boson.
The $W^2$ integration ranges from the pion threshold,
$W_{\pi}^2 = (M + m_\pi)^2$, up to the total center of mass energy
squared, $s = M^2 + 2ME$, while the upper limit on the $Q^2$
integration is $Q^2_{\rm max} = 2ME(1 - W^2/s)$.
Because there is little or no experimental data on the interference
$F_{1,2}^{\gZ}$ structure functions at the kinematics most relevant
to this integral, these need to be estimated from phenomenological
models.

Following Ref.~\citep{Rislow:2010vi}, we divide the integral in
Eq.~(\ref{eq:ImBoxV}) into separate regions in $Q^2$ and $W^2$
according to the dominant physical features and mechanisms
that characterise each region.
  In particular, at low $Q^2$ and $W^2$ 
($Q^2 \leq 10$~GeV$^2$ for $W_\pi^2 \leq W^2 \leq 4$~GeV$^2$
and $Q^2 \leq 2.5$~GeV$^2$ for $4 < W^2 \leq 9$~GeV$^2$, `Region~I'), the structure functions are dominated by nucleon resonances; 
at low $Q^2$ but high $W^2$  
($Q^2 \leq 2.5$~GeV$^2$ and $W^2 > 9$~GeV$^2$, `Region~II') a 
description in terms
of Regge theory is applicable; and at high $Q^2$ and $W^2$ 
($Q^2 > 2.5$~GeV$^2$ and $W^2 > 4$~GeV$^2$, `Region~III')
the deep-inelastic 
structure functions are well described in terms of universal PDFs.

Construction of the $F_{1,2}^{\gZ}$ structure functions requires
firstly choosing appropriate electromagnetic structure functions,
and then transforming these into their $\gZ$ analogs.
In the AJM model, Region~I is well described by the empirical fit to
electron--proton cross section data from Ref.~\cite{Christy:2007ve},
which is quoted with 3--5\% accuracy.
The description of Region~II follows Gorchtein {\it et al.}
\cite{Gorchtein:2011mz} in using the vector meson dominance (VMD)
model together with Regge parametrisations of the high-$W$
behaviour \cite{Alwall:2004wk, Sakurai:1972wk}.
Finally, in Region~III any suitable set of leading twist parton
distributions \cite{JMO13} can be utilised, and in practice we employ
the fit from \mbox{Alekhin {\it et al.}} \cite{Alekhin:2012ig}.

Since the structure functions can be equivalently represented in
terms of the cross sections $\sigma_i$ for the scattering of
transverse ($i=T$) and longitudinal ($i=L$) virtual photons
or $Z$ bosons, it is convenient to separate these into their
resonance and nonresonant background contributions,
  $\sigma_i = \sigma_i^{\rm (res)} + \sigma_i^{\rm (bgd)}$.
These can then be rotated from $\gg \to \gZ$ independently.
For the resonant part, the electromagnetic cross section for the
production of a given resonance $R$ can be modified by the ratio
\cite{Gorchtein:2011mz}
\be
\frac{\sigma_i^{\gZ (R)}}{\sigma_i^{\gg (R)}}\
=\ (1 - 4\sin^2 \theta_W) - y_R,
\ee
where the parameter $y_R$ is computed from the helicity-1/2 and 3/2
nucleon $\to R$ transition amplitudes for the proton and neutron
\cite{Gorchtein:2011mz, Rislow:2010vi, Hall:2013hta}.
While $y_R$ can in principle depend on $Q^2$ in addition to $W^2$,
it was found in Ref.~\cite{Gorchtein:2011mz} that the uncertainty
introduced by approximating $y_R$ to be independent of $Q^2$ is
minimal, and well within the errors on the helicity amplitudes
from the Particle Data Group \cite{Beringer:2012zz}.
The background contribution to the $\gZ$ cross section,
on the other hand, is determined via
  $\sigma_i^{\gZ \rm(bgd)}
   = (\sigma_i^{\gZ}/\sigma_i^{\gg})\, \sigma_i^{\gg \rm (bgd)}$,
where the ratio of the $\gZ$ to $\gg$ cross sections is computed in
the framework of the generalised VMD model \cite{Gorchtein:2011mz},
\be
\frac{\sigma_i^{\gZ}}{\sigma_i^{\gg}}\
=\ \frac{ \kappa_\rho + \kappa_\omega\, R_{\omega}^i(Q^2)
		     + \kappa_\phi\, R_{\phi}^i(Q^2)
		     + \kappa_C^i\, R_C^i(Q^2)}
       {1 + R_{\omega}^i(Q^2)
	  + R_{\phi}^i(Q^2)
	  + R_C^i(Q^2)}.
\label{eq:ghrm45}
\ee
Here the parameters $\kappa_V$ ($V = \rho$, $\omega$, $\phi$) are
ratios of weak and electric charges, while $\kappa_C^i$ denotes the
ratios of the $\gZ$ to $\gg$ continuum contributions.
Similarly, $R_V^i$ are the transverse and longitudinal ratios of
the cross sections for the $V$ and $\rho$ meson, with $R_C^i$
the continuum equivalents.
Although the VMD model does not provide the parameters $\kappa_C^i$,
in the AJM model these were constrained by matching the cross section
ratio in Eq.~(\ref{eq:ghrm45}) with the ratios of PDFs at the boundaries
of Regions~I, II and III.  This was found \cite{Hall:2013hta} to
significantly reduce the uncertainties in $F_{1,2}^{\gZ}$ compared
with earlier estimates \cite{Gorchtein:2011mz}.

As a test of its veracity, the predictions of the AJM model for PV
inelastic asymmetries were compared \cite{Hall:2013hta} with recent
$ed$ scattering data from the E08-011 experiment \cite{Wang:2013kkc} at
Jefferson Lab in the resonance region at $Q^2 = (0.76 - 1.47)$~GeV$^2$,
as well as with earlier $ep$ results from the G0 experiment \cite{G0}
in the $\Delta$ resonance region at $Q^2 = 0.34$~GeV$^2$.
The excellent agreement with the data, which were entirely within the
kinematics defined by Region~I, provides confidence in the extension
of the AJM model to \mo energies.

\begin{figure}[t]
\begin{center}
\includegraphics[width=5in]{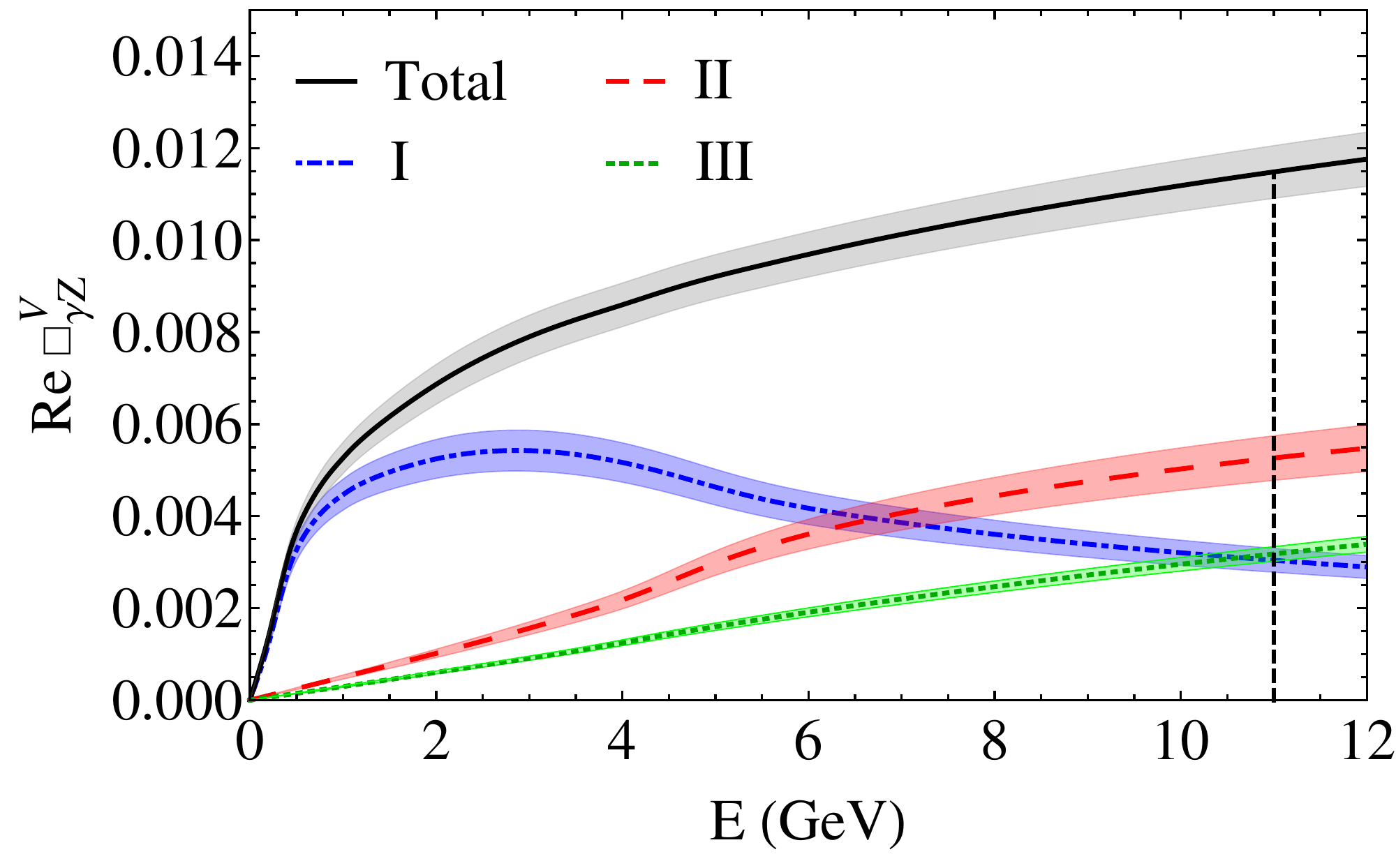}
\caption{(colour online). Energy dependence of the contributions to \regzv\
	from the $W^2$ and $Q^2$ Regions~I (blue dot-dashed line),
	II (red dashed line), and III (green dotted line), together
	with the total (black solid line).  The vertical line
	indicates the energy at the \mo experiment.}
\label{fig:OurReBox}
\end{center}
\end{figure}

\begin{table}[b]
\begin{center} 
\caption{Contributions to \regzv\ in the AJM model from
	Regions~I, II and III at the kinematics of the
	\qwe ($E=1.165$~GeV) and \mo ($E=11$~GeV) experiments.}
\begin{tabular}{ c | c | c }			\hline
	  & \multicolumn{2}{c}{\regzv\ ($\times 10^{-3}$)} \\
Region\ \ & \ \ \qwe \ \ 
	  & \ \ \mo \ \			\\ \hline 
I 	  & $4.64 \pm 0.35$ 
	  & $3.04 \pm 0.26$		\\
II 	  & $0.59 \pm 0.05$ 
	  & $5.26 \pm 0.49$		\\
III 	  & $0.35 \pm 0.02$ 
	  & $3.18 \pm 0.16$		\\ \hline
total 	  & $5.57 \pm 0.36$
	  & $11.5 \pm 0.6$		\\ \hline
\end{tabular}
\label{tab:ReBox}
\end{center}
\end{table}

The energy dependence of the calculated \regzv\ correction in the
AJM model is illustrated in Fig.~\ref{fig:OurReBox}, together with
the contributions from the individual regions, and the values of
the corrections at the \qwe ($E=1.165$~GeV) and \mo ($E=11$~GeV)
energies are listed in Table~\ref{tab:ReBox}.
Compared with the correction relevant for $Q_{\text{weak}}$, the
value of \regzv\ is about twice as large at the \mo energy, where it
is close to one third of the tree level value of $\sin^2\theta_W$.
Furthermore, the relative contributions from the various regions
also change significantly, especially for Region~II, which contributes
only $\approx 11\%$ of the total at the \qwe energy, but yields close
to 50\% of the absolute value at $E=11$~GeV.
Since Regions~I and III contribute less than a third of the total
correction, their roles are relatively less important.

\section{Model dependence}
\label{sec:model}

The AJM model as described above provides our best estimate of the
energy dependence of \regzv\ that is currently possible from existing
experimental constraints.  It is of course important to ensure that
the model dependence of the corrections is accurately reflected in
all relevant uncertainties.  In this section we investigate the
possible model dependence of the most important, Regge contribution
from Region~II, including the dependence on the structure function
parametrisation, and the $W$ and $Q^2$ dependence of the continuum
VMD parameters $\kappa_C^i$.

\begin{figure}[t]
\begin{center}
\includegraphics[width=\textwidth]{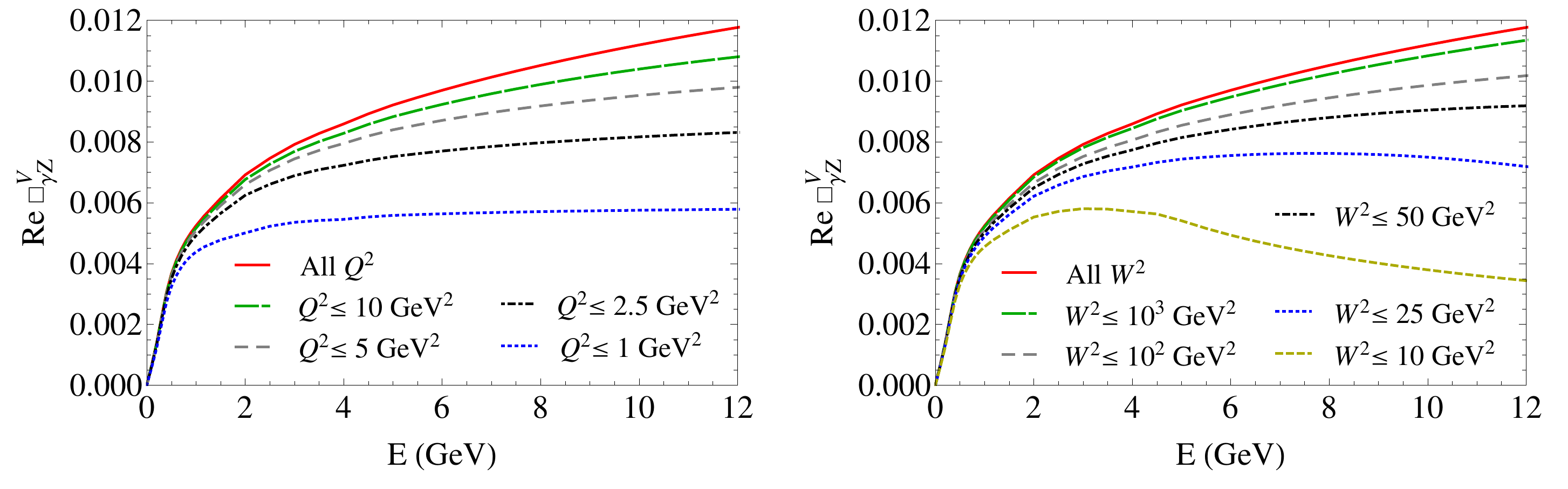}
\caption{(colour online). Contributions to \regzv\ from various kinematic regions
	in $Q^2$ (left) and $W^2$ (right), as a function of the
	energy $E$.}
\label{fig:ReBoxQW}
\end{center}
\end{figure}

To identify the most relevant kinematic regions for the calculation,
we examine the contributions from various $Q^2$ and $W^2$ intervals,
illustrated in Fig.~\ref{fig:ReBoxQW}.  At energies $E \approx 1$~GeV,
the low-$Q^2$ region ($Q^2 < 1$~GeV$^2$) dominates the correction,
while at $E \approx 10$~GeV this region makes up only half of the
total \regzv.  For the $W^2$ intervals, the bulk of the integral
at lower $E$ arises from the low-$W$, nucleon resonance region
($W \lesssim 2-3$~GeV), with larger $W$ becoming more important
with increasing energy.  At $E \approx 10$~GeV, the region
$W^2 > 10$~GeV$^2$ accounts for nearly two thirds of the total.

For the model dependence of the contribution from Region~II,
we consider several different models for the $\gZ$ structure
functions, each of which is based on parametrisations of
electromagnetic cross sections that give reasonable descriptions
of the data at low $W^2$ and high $Q^2$,
but assuming somewhat different physical mechanisms for the
scattering in this region.
For example, the Regge theory inspired parametrisation of
Capella {\it et al.} \cite{Capella:1994cr} was used in the
calculation by Sibirtsev {\it et al.} \cite{Sibirtsev:2010zg},
while the fit of Alwall and Ingleman \cite{Alwall:2004wk},
based on a combination of the VMD model and Regge phenomenology,
was used as the basis of `Model~II' of Gorchtein {\it et al.}
\cite{Gorchtein:2011mz}.  The colour dipole model (CDP) of Cvetic
{\it et al.} \cite{Cvetic:1999fi, Cvetic:2001ie} was also used in
Ref.~\cite{Gorchtein:2011mz} in what was referred to as `Model~I'.
Note that while `Model~I' considered photon couplings at the
hadronic level, `Model~II' assumed couplings to quarks directly.

More specifically, in addition to the AJM $\gZ$ structure function
model, we compare the results for \regzv\ with several alternative
models:

\subsubsection*{Modified Regge model (MRM)}
As in the original calculation of Ref.~\cite{Sibirtsev:2010zg},
the MRM uses the Capella {\it et al.} parametrisation
\cite{Capella:1994cr} of $F_{1,2}^{\gg}$, although instead of using
leading twist PDFs to modify the structure functions, here the
$\gg \to \gZ$ rotation is performed via Eq.~(\ref{eq:ghrm45}).
Since this parametrisation is not separated into resonance and
nonresonance components, the entire $F_{1,2}^{\gg}$ structure
functions must be rotated.  This may appear {\it ad hoc}
(since the resonances are all scaled by the same amount),
however, because the resonance contribution is negligible in
Region~II, the total transverse and longitudinal cross sections
$\sigma_i$ are effectively given by their background contributions,
$\sigma_i \approx \sigma_i^{\rm (bgd)}$.

\subsubsection*{CDP model}  
The CDP parametrisation of the electromagnetic structure functions 
\cite{Cvetic:1999fi, Cvetic:2001ie} formed the basis of `Model~I'
in Ref.~\cite{Gorchtein:2011mz}.  Instead of using the VMD inspired
relation in Eq.~(\ref{eq:ghrm45}), however, the $\kappa_V$ and
$\kappa_C^i$ coefficients were computed from ratios of quark electric
charges.  The resulting ratio of $\gZ$ to $\gg$ cross sections in this
model is then given by a constant value \cite{Gorchtein:2011mz},
\be
\frac{\sigma_i^{\gZ}}
     {\sigma_i^{\gg}}\
=\ \frac{9}{5} - 4 \sin^2 \theta_W.
%
\label{eq:ghrm52}
\ee
In the implementation of the CDP model in the present analysis,
an updated parametrisation of the $\gg$ cross section
\cite{Kuroda:2011dw} is used.

\subsubsection*{CDP/VMD model} 
This model combines the electromagnetic structure functions of Cvetic
{\it et al.} \cite{Cvetic:1999fi, Cvetic:2001ie} with the constrained
$\gZ/\gg$ ratio in Eq.~(\ref{eq:ghrm45}), as in the AJM model.
Note that for both the CDP and CDP/VMD models the parametrisation
\cite{Cvetic:1999fi, Cvetic:2001ie} of the structure functions is
given only for $W^2 < 1000$~GeV$^2$.  
As shown in Fig.~\ref{fig:ReBoxQW}, the contribution to the integral
from $W^2>1000\gev^2$ is only a very small fraction of the total
correction to \regzv.  For the CDP and CDP/VMD models, we estimate
the fractional contribution to the integral from the $W^2>1000\gev^2$
region to be the same as in the AJM model.

\vspace*{0.5cm}

\begin{figure}[t]
\begin{center}
\includegraphics[width=5in]{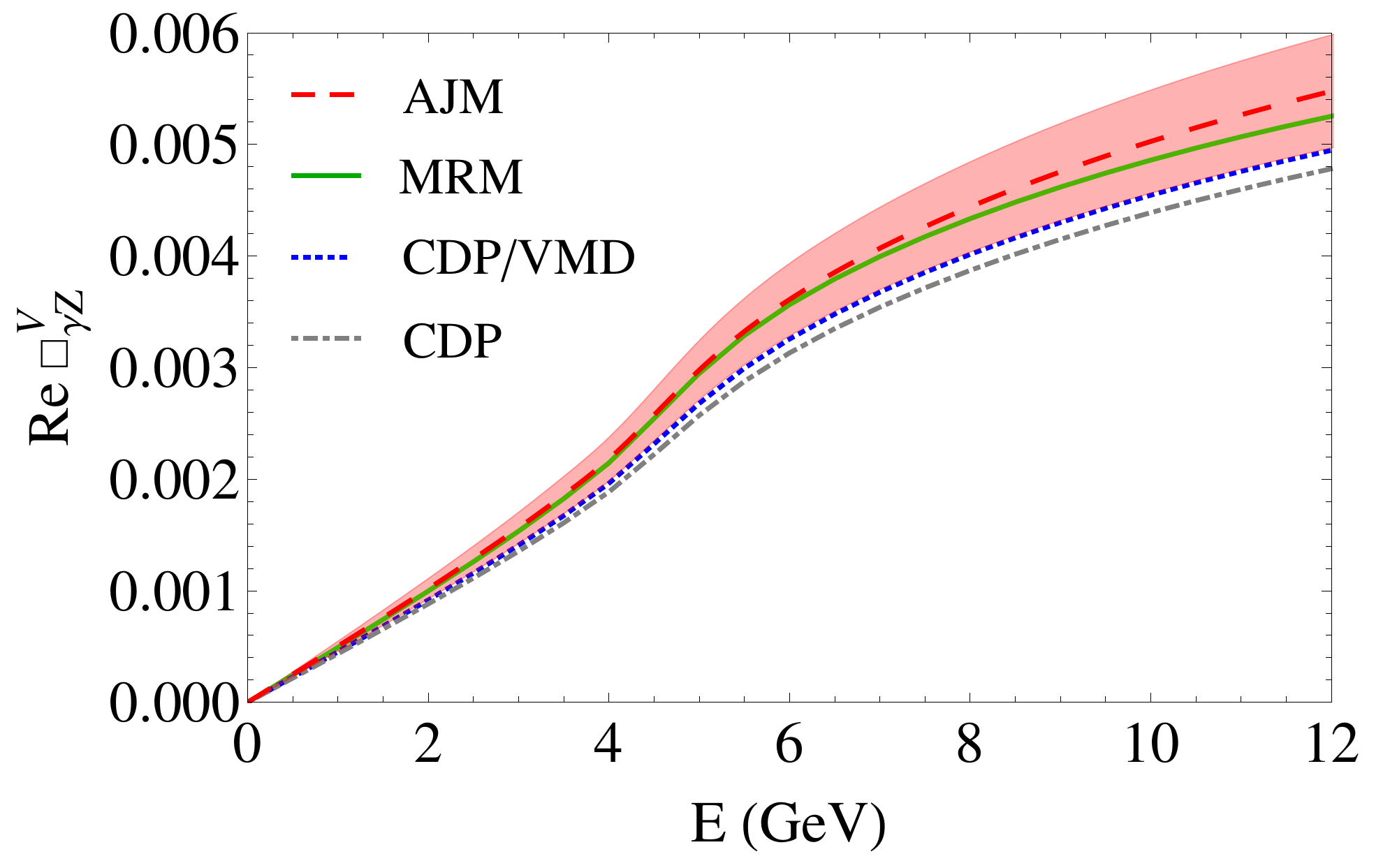}
\caption{(colour online). Contribution of Region~II to \regzv as a function of energy
	using various models for the $\gZ$ structure functions:
	AJM (red dashed line),
	MRM (green solid line),
	CDP/VDM (blue dotted line), and
	CDP (gray dot-dashed line).}
\label{fig:ReIIMod}
\end{center}
\end{figure}

Using these alternative models for the $\gZ$ interference structure
functions,
which assume rather different physical pictures,
the contributions from Region~II to \regzv\ are shown in
Fig.~\ref{fig:ReIIMod}, compared with the results of the AJM model
of Ref.~\cite{Hall:2013hta}.  The central values of the corrections
using the AJM and MRM $\gZ$ structure functions are very similar over
the entire range of energies considered, while the CDP and CDP/VMD
models give slightly smaller corrections.
The uncertainty band of AJM model includes the MRM and CDP/VMD
results, with the CDP lying slightly below.
To ensure that the overall \regzv\ error is not underestimated,
we include the full difference between the most disparate central
values, combining it in quadrature with the AJM estimate of the error,
as an additional uncertainty arising from the model dependence.

A further error on \regzv\ could arise from the dependence of the
continuum parameters $\kappa_C^i$ on the invariant hadronic mass $W$.
In the AJM model, the continuum coefficients were fitted by constant
values
  $\kappa_C^T = 0.65 \pm 0.14$ and
  $\kappa_C^L= -1.3 \pm 1.7$
\cite{Hall:2013hta},
with the possible $W^2$ dependence taken into account by appropriate
matching over the range between $W^2=4$~GeV$^2$ and 13~GeV$^2$, and
including any variation in the uncertainties.  Increasing the range over
which these are fitted to $4 \leq W^2 \leq 1000$~GeV$^2$, one finds
  $\kappa_C^T = 0.86 \pm 0.24$ and
  $\kappa_C^L= -1.3 \pm 2.3$,
which are consistent with the values obtained in
Ref.~\cite{Hall:2013hta} within the uncertainties.
While there is some variation of the resulting $\gZ$ structure functions
computed from the two sets of values, the uncertainties assigned to the
$\kappa_C^i$ values in the AJM model are sufficient to cover the different
behaviours.  Since the additional uncertainty is found to be negligible,
we do not assign an additional error on \regzv\ from the $W^2$ dependence
of the $\kappa_C^i$.

As well as the $W^2$ dependence, we also consider the $Q^2$ dependence
of the $\kappa_C^i$ errors in the region $0 \leq Q^2 \leq 2.5$~GeV$^2$.
Taking the AJM model errors for $\kappa_C^i$ at $Q^2 = 2.5$~GeV$^2$,
where the constraints from PDFs are expected to be reliable, the error
is increased linearly to 100\% at the real photon point, $Q^2 = 0$.
With these modified constraints the uncertainty on \regzv\ would
increase from
  $\pm\, 0.36 \times 10^{-3}$ to
  $\pm\, 0.59 \times 10^{-3}$ at the \qwe energy,
while for \mo the error would double to $\pm\, 1.2 \times 10^{-3}$.
In either case, these errors would still remain within the
experimental budget; in practice, however, we believe they are
likely to be too conservative and take the error on $\kappa_C^T$
to be constant in $Q^2$.

Using the AJM model with constraints provided by PDFs, the relative 
contributions to \regzv\ relevant to the \mo experiment from the various
kinematic regions differ significantly from the those at the \qwe
energy.  In particular, the contribution from Region~II is much larger 
(by $\sim 50\%$) than that at lower energies.  
Taking into account the additional model dependence discussed in
this section, the final value for \regzv\ at the \mo energy is
estimated to be
\be
\Re e\, \square_{\gZ}^V
= (11.5 \pm 0.6 \pm 0.6) \times 10^{-3},
\label{eq:ReBox_final}
\ee
where the first error includes the various sources of uncertainty
in the AJM model, while the second arises from the additional model
dependence considered in this analysis.  Further experiments to
determine the $\gZ$ interference structure functions would naturally
increase the precision of this result.
Nevertheless, with the current precision of \regzv, the effective
weak charge of the proton increases from
  $0.0757 \pm 0.0007$ at $E=1.165$~GeV to
  $0.0814 \pm 0.0010$ at $E=11$~GeV. 
Included in this effective $Q_W^p$ is the contribution from the hadronic
axial-vector piece (with leptonic vector coupling to the $Z$), \bgZa.
By extending the work of Refs.~\cite{Blunden:2011rd, Blunden:2012ty},
\bgZa decreases from 
  0.0037(2) at $E=1.165$~GeV to 
  0.0035(2) at $E=11$~GeV.
Since PV $ep$ elastic scattering is estimated to constitute a background
of the order 8\% to the M\o ller measurement, the uncertainty it will
induce is $\approx 0.1\%$.  This figure is well below the limit of 0.3\%
anticipated from elastic proton, as reported in the experimental proposal.

\section{Inelastic asymmetry}

For the \mo experiment, the inelastic $ep$ cross section is an order
of magnitude smaller than the elastic background, yet the inelastic
$Z$ coupling to the proton is not suppressed by the proton weak charge
$Q_W^p$. This increases the relative significance of this inelastic 
contribution to the asymmetry, and therefore important to consider.

Within the dispersion relation approach, the main uncertainty in the
calculation of \regzv\ is the $\gZ$ interference structure functions.
In principle these can be measured directly in parity-violating
deep-inelastic scattering.  In fact, the inelastic PV asymmetry
involves the same combination of $F_{1,2}^{\gZ}$ as that appearing
in the integrand of Eq.~(\ref{eq:ImBoxV}), and has been used to
constrain the $\gZ$ box corrections.
As discussed above, the AJM model makes full use of the available data
on PV inelastic $ep$ \cite{G0} and $ed$ \cite{Wang:2013kkc} scattering
at low and intermediate $Q^2$ values, in addition to constraints from
parton distributions at high $W$ and $Q^2$.

\begin{figure}[t]
\begin{center}
\includegraphics[width=5in]{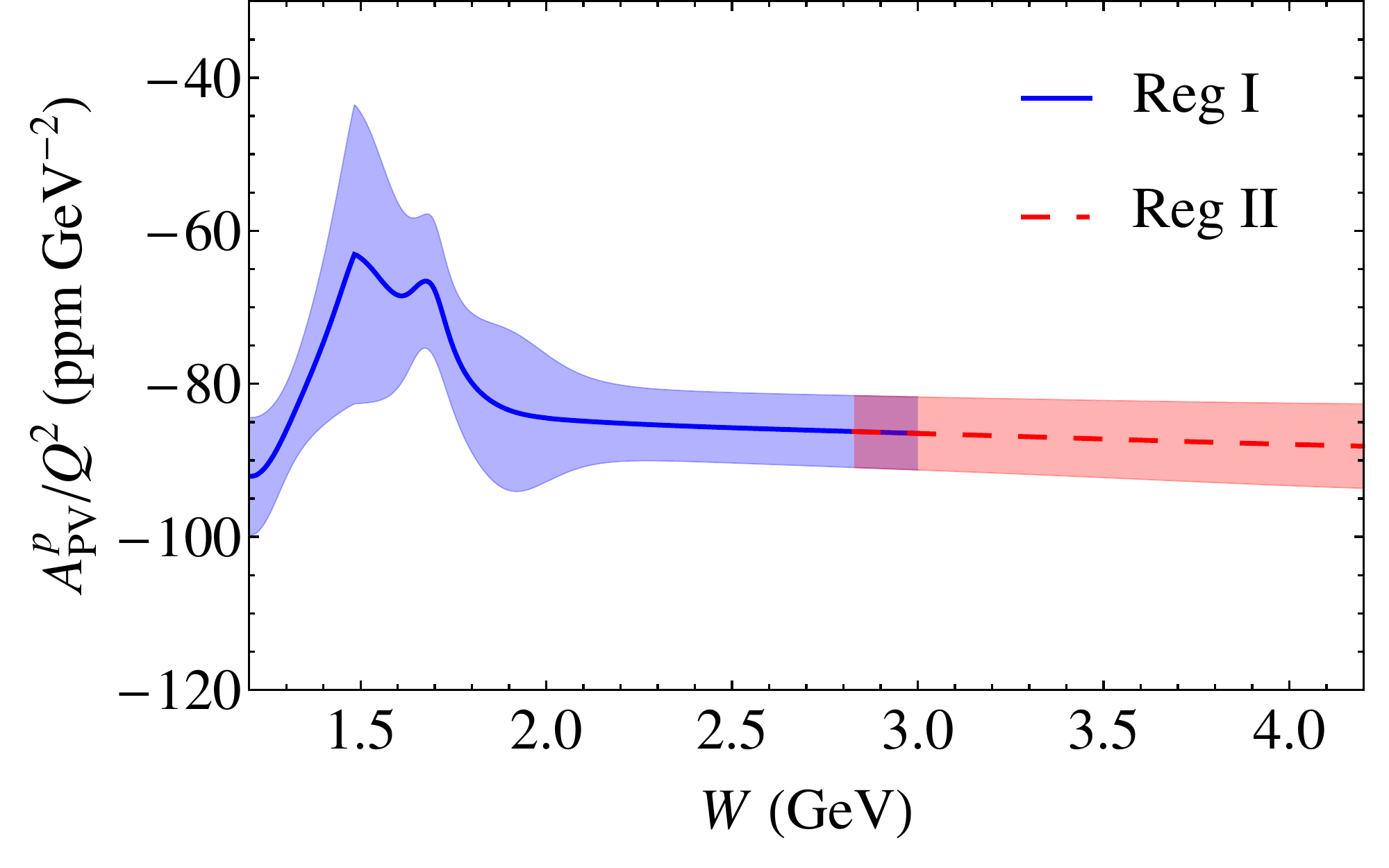}
\caption{(colour online). PV asymmetry $A_{\rm PV}^p$ for inelastic $ep$ scattering,
	scaled by $1/Q^2$, at $E=11$~GeV in the AJM model, showing the
	matching of the contributions from Region~I (blue solid line)
	and Region~II (red dashed line).  The typical momentum transfer
	relevant to the \mo experiment is $\approx 0.004$~GeV$^2$.}
\label{fig:inel}
\end{center}
\end{figure}

The PV inelastic asymmetry for $ep$ scattering is shown in
Fig.~\ref{fig:inel} in the AJM model for kinematics typical
for the \mo experiment, for which $Q^2 \approx 0.004$~GeV$^2$.
At low $W$ ($W \lesssim 2$~GeV) the asymmetry illustrates the
structure characteristic of the nucleon resonance region,
while at higher $W$ the asymmetry (scaled by $1/Q^2$) remains
approximately constant at $\approx 85$~ppm/GeV$^2$, with an
uncertainty of $\approx 7\%$ in the AJM model.

As discussed in the previous section, a more conservative error
estimate could amplify the uncertainties on the $\kappa_C^i$ to 100\%
at the photon point.  In this case, the relative uncertainty on the
inelastic PV asymmetry in the \mo experiment would increase to about
25\%.  It will be important, therefore, to have additional PV inelastic
data to further constrain the model, and to emprically constrain the
inelastic background in the \mo experiment itself.
Such direct monitoring of the inelastic $ep$ asymmetry was achieved
in the E158 experiment at SLAC, where this background was resolved at
better than 20\% precision \cite{Anthony:2005pm}.

Note also that in this estimate other standard radiative corrections
to PV inelastic scattering have not been included.  These may modify
the specific $W$ distribution shown in Fig.~\ref{fig:inel}; however,
for the purposes of illustrating the behaviour of the asymmetry and its
uncertainty our estimate is sufficient.

\section{Conclusion}

In summary, we have calculated the energy dependent $\gZ$ radiative
correction to PV elastic $ep$ scattering to an accuracy of
$\approx 7\%$, at an energy of $E=11$~GeV relevant for the planned
\mo experiment at \mbox{Jefferson Lab}.  In contrast to the \qwe
experiment kinematics, where the resonance region dominates, the
energy-dependent \regzv\ correction at \mo kinematics is much more
sensitive to the Regge region.
With careful attention paid to the model dependence of the $\gZ$
structure functions in the Regge region, we determine an effective
proton weak charge of $0.0814 \pm 0.0010$ at 11~GeV.  This represents
a precision that is sufficient to keep the uncertainty from this
background within the limits of the \mo experiment.

We have also used the AJM model to estimate the magnitude and shape
of the inelastic $ep$ scattering background. Although the AJM model
sufficiently constrains the magnitude of this background, a more
conservative estimate emphasises the importance of directly monitoring
this background within the \mo experiment.  Conversely, this measurement
may serve to better test or constrain the inputs of the AJM model.

\section*{Acknowledgements}

We thank K.~Kumar for helpful discussions and M.~Kuroda for providing
code of the colour dipole model parametrisation.  N.H. and P.B. thank
the Jefferson Lab Theory Center, and W.M. thanks the CSSM/CoEPP at
the University of Adelaide and the University of Manitoba for support
during visits where some of this work was performed.  This work was
supported by NSERC (Canada), the DOE Contract No. DE-AC05-06OR23177,
under which Jefferson Science Associates, LLC operates Jefferson Lab,
and the Australian Research Council through an Australian Laureate
Fellowship (A.W.T.), a Future Fellowship (R.D.Y.) and through the
ARC Centre of Excellence for Particle Physics at the Terascale.

%
%
\bibliographystyle{model1-num-names}
\bibliography{Mollbib}

\begin{thebibliography}{39}
\expandafter\ifx\csname natexlab\endcsname\relax\def\natexlab#1{#1}\fi
\providecommand{\url}[1]{\texttt{#1}}
\providecommand{\href}[2]{#2}
\providecommand{\path}[1]{#1}
\providecommand{\DOIprefix}{doi:}
\providecommand{\ArXivprefix}{arXiv:}
\providecommand{\URLprefix}{URL: }
\providecommand{\Pubmedprefix}{pmid:}
\providecommand{\doi}[1]{\href{http://dx.doi.org/#1}{\path{#1}}}
\providecommand{\Pubmed}[1]{\href{pmid:#1}{\path{#1}}}
\providecommand{\bibinfo}[2]{#2}
\ifx\xfnm\relax \def\xfnm[#1]{\unskip,\space#1}\fi
\bibitem[{Prescott et~al.(1979)}]{Prescott:1979dh}
\bibinfo{author}{C.~Y. Prescott}, et~al.,
\newblock \bibinfo{journal}{Phys. Lett. B} \bibinfo{volume}{84}
  (\bibinfo{year}{1979}) \bibinfo{pages}{524}.
\bibitem[{Souder et~al.(1990)}]{Souder:1990ia}
\bibinfo{author}{P.~A. Souder}, et~al.,
\newblock \bibinfo{journal}{Phys. Rev. Lett.} \bibinfo{volume}{65}
  (\bibinfo{year}{1990}) \bibinfo{pages}{694}.
\bibitem[{Aniol et~al.(2001)}]{Aniol:2000at}
\bibinfo{author}{K.~A. Aniol}, et~al.,
\newblock \bibinfo{journal}{Phys. Lett. B} \bibinfo{volume}{509}
  (\bibinfo{year}{2001}) \bibinfo{pages}{211}.
\bibitem[{Aniol et~al.(2006)}]{Aniol:2005zg}
\bibinfo{author}{K.~A. Aniol}, et~al.,
\newblock \bibinfo{journal}{Phys. Lett. B} \bibinfo{volume}{635}
  (\bibinfo{year}{2006}) \bibinfo{pages}{275}.
\bibitem[{Armstrong et~al.(2005)}]{Armstrong:2005hs}
\bibinfo{author}{D.~S. Armstrong}, et~al.,
\newblock \bibinfo{journal}{Phys. Rev. Lett.} \bibinfo{volume}{95}
  (\bibinfo{year}{2005}) \bibinfo{pages}{092001}.
\bibitem[{Paschke et~al.(2011)Paschke, Thomas, Michaels, and
  Armstrong}]{Paschke:2011zz}
\bibinfo{author}{K.~Paschke}, \bibinfo{author}{A.~W. Thomas},
  \bibinfo{author}{R.~Michaels}, \bibinfo{author}{D.~Armstrong},
\newblock \bibinfo{journal}{J. Phys. Conf. Ser.} \bibinfo{volume}{299}
  (\bibinfo{year}{2011}) \bibinfo{pages}{012003}.
\bibitem[{Young et~al.(2006)Young, Roche, Carlini, and Thomas}]{Young:2006jc}
\bibinfo{author}{R.~D. Young}, \bibinfo{author}{J.~Roche},
  \bibinfo{author}{R.~D. Carlini}, \bibinfo{author}{A.~W. Thomas},
\newblock \bibinfo{journal}{Phys. Rev. Lett.} \bibinfo{volume}{97}
  (\bibinfo{year}{2006}) \bibinfo{pages}{102002}.
\bibitem[{Young et~al.(2007)Young, Carlini, Thomas, and Roche}]{Young:2007zs}
\bibinfo{author}{R.~D. Young}, \bibinfo{author}{R.~D. Carlini},
  \bibinfo{author}{A.~W. Thomas}, \bibinfo{author}{J.~Roche},
\newblock \bibinfo{journal}{Phys. Rev. Lett.} \bibinfo{volume}{99}
  (\bibinfo{year}{2007}) \bibinfo{pages}{122003}.
\bibitem[{Roche et~al.(2011)Roche, van Oers, and Young}]{Roche:2011zz}
\bibinfo{author}{J.~Roche}, \bibinfo{author}{W.~T.~H. van Oers},
  \bibinfo{author}{R.~D. Young},
\newblock \bibinfo{journal}{J. Phys. Conf. Ser.} \bibinfo{volume}{299}
  (\bibinfo{year}{2011}) \bibinfo{pages}{012012}.
\bibitem[{Androic et~al.(2013)}]{Qweak13}
\bibinfo{author}{D.~Androic}, et~al.,
\newblock \bibinfo{journal}{Phys. Rev. Lett.} \bibinfo{volume}{111}
  (\bibinfo{year}{2013}) \bibinfo{pages}{141803}.
\bibitem[{Wang et~al.(2013)}]{Wang:2013kkc}
\bibinfo{author}{D.~Wang}, et~al.,
\newblock \bibinfo{journal}{Phys. Rev. Lett.} \bibinfo{volume}{111}
  (\bibinfo{year}{2013}) \bibinfo{pages}{082501}.
\bibitem[{Mammei(2012)}]{Mammei:2012ph}
\bibinfo{author}{J.~Mammei},
\newblock \bibinfo{journal}{Nuovo Cim.} \bibinfo{volume}{C035N04}
  (\bibinfo{year}{2012}) \bibinfo{pages}{203}.
\bibitem[{Souder(2012)}]{Souder:2012zz}
\bibinfo{author}{P.~A. Souder},
\newblock \bibinfo{journal}{AIP Conf. Proc.} \bibinfo{volume}{1441}
  (\bibinfo{year}{2012}) \bibinfo{pages}{123}.
\bibitem[{Erler et~al.(2003)Erler, Kurylov, and Ramsey-Musolf}]{Erler:2003yk}
\bibinfo{author}{J.~Erler}, \bibinfo{author}{A.~Kurylov},
  \bibinfo{author}{M.~J. Ramsey-Musolf},
\newblock \bibinfo{journal}{Phys. Rev. D} \bibinfo{volume}{68}
  (\bibinfo{year}{2003}) \bibinfo{pages}{016006}.
\bibitem[{Erler and Su(2013)}]{Erler:2013xha}
\bibinfo{author}{J.~Erler}, \bibinfo{author}{S.~Su},
\newblock \bibinfo{journal}{Prog. Part. Nucl. Phys.} \bibinfo{volume}{71}
  (\bibinfo{year}{2013}) \bibinfo{pages}{119}.
\bibitem[{Derman and Marciano(1979)}]{Derman:1979zc}
\bibinfo{author}{E.~Derman}, \bibinfo{author}{W.~J. Marciano},
\newblock \bibinfo{journal}{Annals Phys.} \bibinfo{volume}{121}
  (\bibinfo{year}{1979}) \bibinfo{pages}{147}.
\bibitem[{Czarnecki and Marciano(1996)}]{Czarnecki:1995fw}
\bibinfo{author}{A.~Czarnecki}, \bibinfo{author}{W.~J. Marciano},
\newblock \bibinfo{journal}{Phys. Rev. D} \bibinfo{volume}{53}
  (\bibinfo{year}{1996}) \bibinfo{pages}{1066}.
\bibitem[{Denner and Pozzorini(1999)}]{Denner:1998um}
\bibinfo{author}{A.~Denner}, \bibinfo{author}{S.~Pozzorini},
\newblock \bibinfo{journal}{Eur. Phys. J. C} \bibinfo{volume}{7}
  (\bibinfo{year}{1999}) \bibinfo{pages}{185}.
\bibitem[{Petriello(2003)}]{Petriello:2002wk}
\bibinfo{author}{F.~J. Petriello},
\newblock \bibinfo{journal}{Phys. Rev. D} \bibinfo{volume}{67}
  (\bibinfo{year}{2003}) \bibinfo{pages}{033006}.
\bibitem[{Gorchtein and Horowitz(2009)}]{Gorchtein:2008px}
\bibinfo{author}{M.~Gorchtein}, \bibinfo{author}{C.~J. Horowitz},
\newblock \bibinfo{journal}{Phys. Rev. Lett.} \bibinfo{volume}{102}
  (\bibinfo{year}{2009}) \bibinfo{pages}{091806}.
\bibitem[{Sibirtsev et~al.(2010)Sibirtsev, Blunden, Melnitchouk, and
  Thomas}]{Sibirtsev:2010zg}
\bibinfo{author}{A.~Sibirtsev}, \bibinfo{author}{P.~G. Blunden},
  \bibinfo{author}{W.~Melnitchouk}, \bibinfo{author}{A.~W. Thomas},
\newblock \bibinfo{journal}{Phys. Rev. D} \bibinfo{volume}{82}
  (\bibinfo{year}{2010}) \bibinfo{pages}{013011}.
\bibitem[{Rislow and Carlson(2011)}]{Rislow:2010vi}
\bibinfo{author}{B.~C. Rislow}, \bibinfo{author}{C.~E. Carlson},
\newblock \bibinfo{journal}{Phys. Rev. D} \bibinfo{volume}{83}
  (\bibinfo{year}{2011}) \bibinfo{pages}{113007}.
\bibitem[{Gorchtein et~al.(2011)Gorchtein, Horowitz, and
  Ramsey-Musolf}]{Gorchtein:2011mz}
\bibinfo{author}{M.~Gorchtein}, \bibinfo{author}{C.~J. Horowitz},
  \bibinfo{author}{M.~J. Ramsey-Musolf},
\newblock \bibinfo{journal}{Phys. Rev. C} \bibinfo{volume}{84}
  (\bibinfo{year}{2011}) \bibinfo{pages}{015502}.
\bibitem[{Blunden et~al.(2011)Blunden, Melnitchouk, and
  Thomas}]{Blunden:2011rd}
\bibinfo{author}{P.~G. Blunden}, \bibinfo{author}{W.~Melnitchouk},
  \bibinfo{author}{A.~W. Thomas},
\newblock \bibinfo{journal}{Phys. Rev. Lett.} \bibinfo{volume}{107}
  (\bibinfo{year}{2011}) \bibinfo{pages}{081801}.
\bibitem[{Hall et~al.(2013)Hall, Blunden, Melnitchouk, Thomas, and
  Young}]{Hall:2013hta}
\bibinfo{author}{N.~L. Hall}, \bibinfo{author}{P.~G. Blunden},
  \bibinfo{author}{W.~Melnitchouk}, \bibinfo{author}{A.~W. Thomas},
  \bibinfo{author}{R.~D. Young},
\newblock \bibinfo{journal}{Phys. Rev. D} \bibinfo{volume}{88}
  (\bibinfo{year}{2013}) \bibinfo{pages}{013011}.
\bibitem[{Arrington et~al.(2011)Arrington, Blunden, and
  Melnitchouk}]{Arrington:2011dn}
\bibinfo{author}{J.~Arrington}, \bibinfo{author}{P.~G. Blunden},
  \bibinfo{author}{W.~Melnitchouk},
\newblock \bibinfo{journal}{Prog. Part. Nucl. Phys.} \bibinfo{volume}{66}
  (\bibinfo{year}{2011}) \bibinfo{pages}{782}.
\bibitem[{Christy and Bosted(2010)}]{Christy:2007ve}
\bibinfo{author}{M.~E. Christy}, \bibinfo{author}{P.~E. Bosted},
\newblock \bibinfo{journal}{Phys. Rev. C} \bibinfo{volume}{81}
  (\bibinfo{year}{2010}) \bibinfo{pages}{055213}.
\bibitem[{Alwall and Ingelman(2004)}]{Alwall:2004wk}
\bibinfo{author}{J.~Alwall}, \bibinfo{author}{G.~Ingelman},
\newblock \bibinfo{journal}{Phys. Lett. B} \bibinfo{volume}{596}
  (\bibinfo{year}{2004}) \bibinfo{pages}{77}.
\bibitem[{Sakurai and Schildknecht(1972)}]{Sakurai:1972wk}
\bibinfo{author}{J.~J. Sakurai}, \bibinfo{author}{D.~Schildknecht},
\newblock \bibinfo{journal}{Phys. Lett. B} \bibinfo{volume}{40}
  (\bibinfo{year}{1972}) \bibinfo{pages}{121}.
\bibitem[{Jimenez-Delgado et~al.(2013)Jimenez-Delgado, Melnitchouk, and
  Owens}]{JMO13}
\bibinfo{author}{P.~Jimenez-Delgado}, \bibinfo{author}{W.~Melnitchouk},
  \bibinfo{author}{J.~F. Owens},
\newblock \bibinfo{journal}{J. Phys. G} \bibinfo{volume}{40}
  (\bibinfo{year}{2013}) \bibinfo{pages}{093102}.
\bibitem[{Alekhin et~al.(2012)Alekhin, Blumlein, and Moch}]{Alekhin:2012ig}
\bibinfo{author}{S.~Alekhin}, \bibinfo{author}{J.~Blumlein},
  \bibinfo{author}{S.~Moch},
\newblock \bibinfo{journal}{Phys. Rev. D} \bibinfo{volume}{86}
  (\bibinfo{year}{2012}) \bibinfo{pages}{054009}.
\bibitem[{Beringer et~al.(2012)}]{Beringer:2012zz}
\bibinfo{author}{J.~Beringer}, et~al.,
\newblock \bibinfo{journal}{Phys. Rev. D} \bibinfo{volume}{86}
  (\bibinfo{year}{2012}) \bibinfo{pages}{010001}.
\bibitem[{Androic et~al.(2012)}]{G0}
\bibinfo{author}{D.~Androic}, et~al.,
\newblock \bibinfo{journal}{arXiv:1212.1637}  (\bibinfo{year}{2012}).
\bibitem[{Capella et~al.(1994)Capella, Kaidalov, Merino, and Tran
  Thanh~Van}]{Capella:1994cr}
\bibinfo{author}{A.~Capella}, \bibinfo{author}{A.~Kaidalov},
  \bibinfo{author}{C.~Merino}, \bibinfo{author}{J.~Tran Thanh~Van},
\newblock \bibinfo{journal}{Phys. Lett. B} \bibinfo{volume}{337}
  (\bibinfo{year}{1994}) \bibinfo{pages}{358}.
\bibitem[{Cvetic et~al.(2000)Cvetic, Schildknecht, and Shoshi}]{Cvetic:1999fi}
\bibinfo{author}{G.~Cvetic}, \bibinfo{author}{D.~Schildknecht},
  \bibinfo{author}{A.~Shoshi},
\newblock \bibinfo{journal}{Eur. Phys. J. C} \bibinfo{volume}{13}
  (\bibinfo{year}{2000}) \bibinfo{pages}{301}.
\bibitem[{Cvetic et~al.(2001)Cvetic, Schildknecht, Surrow, and
  Tentyukov}]{Cvetic:2001ie}
\bibinfo{author}{G.~Cvetic}, \bibinfo{author}{D.~Schildknecht},
  \bibinfo{author}{B.~Surrow}, \bibinfo{author}{M.~Tentyukov},
\newblock \bibinfo{journal}{Eur. Phys. J. C} \bibinfo{volume}{20}
  (\bibinfo{year}{2001}) \bibinfo{pages}{77}.
\bibitem[{Kuroda and Schildknecht(2012)}]{Kuroda:2011dw}
\bibinfo{author}{M.~Kuroda}, \bibinfo{author}{D.~Schildknecht},
\newblock \bibinfo{journal}{Phys. Rev. D} \bibinfo{volume}{85}
  (\bibinfo{year}{2012}) \bibinfo{pages}{094001}.
\bibitem[{Blunden et~al.(2012)Blunden, Melnitchouk, and
  Thomas}]{Blunden:2012ty}
\bibinfo{author}{P.~G. Blunden}, \bibinfo{author}{W.~Melnitchouk},
  \bibinfo{author}{A.~W. Thomas},
\newblock \bibinfo{journal}{Phys. Rev. Lett.} \bibinfo{volume}{109}
  (\bibinfo{year}{2012}) \bibinfo{pages}{262301}.
\bibitem[{Anthony et~al.(2005)}]{Anthony:2005pm}
\bibinfo{author}{P.~L. Anthony}, et~al.,
\newblock \bibinfo{journal}{Phys. Rev. Lett.} \bibinfo{volume}{95}
  (\bibinfo{year}{2005}) \bibinfo{pages}{081601}.

\end{thebibliography}

\end{document}